\documentstyle[aps]{revtex}

\begin{document}
\draft

\title{analytical calculation of the Peierls-Nabarro pinning barrier for one-dimensional parametric double-well models}
\author{Alain M. Dikand\'e}
\address{Centre de Recherches sur les Propri\'et\'es \'Electroniques des Mat\'eriaux Avanc\'es, D\'epartement de 
Physique, Facult\'e des Sciences, Universit\'e de Sherbrooke J1K-2R1 Sherbrooke-Qu\'ebec, CANADA \\ 
electronic mail: amdikand@physique.usherb.ca}

\date{\today}

\maketitle

\begin{abstract}
Lattice effects on the kink families of two models for one-dimensional nonlinear Klein-Gordon 
systems with double-well on-site potentials are considered. The analytical 
expression of the generalized Peierls-Nabarro pinning potential is obtained and confronted with numerical simulations. 
\end{abstract}

\pacs{}


Many physical problems have greatly contributed to the interest on
double-well potential(DWP) models~\cite{remoissenet}. Among the
most important ones are the mechanism of protonic conductivity in
molecular systems~\cite{kuwada,davidov}, structural instabilities
in one-dimensional(1D) Perovskyte-like systems~\cite{varma} as
well as Hydrogen-bonded ferroelectrics \cite{blinc},  and  Peierls
transition in 1D conductors~\cite{heeger}. The first problem ows
its interest to the fundamental role of proton migrations among
the surrounding(heavy) ions of the 1D molecular system. The 1D
character of these systems sits in the very high conductivity
along the Hydrogen-Bond(HB)-about $10^{3}$ times greater-compared
to the two other directions~\cite{vanderkoy}. This strong
anisotropic conductivity relies on structural defects, that is
ionic and bonding defects~\cite{bernal}. Defects are consequent
upon proton hopping and bond distortions across the energy barrier
between the two most stretched proton positions along the HB. When
evoking ferroelectricity and the Peierls transition(an instability
of the electronic structure of a 1D electron-phonon system), the
most natural framework is that of the Landau theory of
second-order phase transition~\cite{landau}. The "DWP" scheme in
this theory is appropriate for such phenomena as the proton
ordering transitions in the pseudo-1D ferroelectrics
$KH_{2}PO_{4}$~\cite{blinc} and structural instabilities in 1D
organic and inorganic conductors~\cite{heeger}. For all these
systems, in an idealized context in the absence of external
bias(electric or magnetic fields), defect nucleations(soliton
lattice)~\cite{jensen} and the intrinsic discrete structure of the
background substrate lattice~\cite{aubry} are the two main
processes influencing conductivity. In general they give rise to
an activated transport property that consists either in a
conducting phase or in an insulating phase~\cite{bak1,pokrovsky}.
In concrete terms, by the Kramers theory of diffusion it is easy
to point out that the kink diffusion coefficient in a discrete
medium will be weighted by an Arhenius-like factor $R(\omega_p)$ which
accounts for the escape rate of kink from the lattice trapping.
This factor depends on the vibration frequency $\omega_p$ of the
kink in its attempts to jump across the pinning potential barrier,
thereby determining its lifetime in the discrete medium. At low
temperatures, the discrete kink diffusion coefficient is of the
form
\begin{equation}
D \sim D_o R(\omega_p), \hspace{.2in} R(\omega_p) \sim \omega_p \exp(-\beta E_{PN})  \label{int}
\end{equation}
 $E_{PN}$ in this formula is the lattice-induced trapping 
potential barrier, the kink pinning
frequency $\omega_p$ is directly connected to this physical
quantity. The activated diffusion coefficient given above is not
only interesting in its own right, but also it furnishes relevant
knowledge for capturing the essential features of all other
transport parameters namely, the kink mobility and conductivity. \\
 While DWP models have so become useful in the understanding of
physical problems, insights gained from the earliest model i.e.
the ${\phi}^{4}$ rapidly saturated owing to several shortcomings.
For instance, the rigidity prevents from getting further in the
theory as to analytically account for the possibility of shifting
the potential minima and hump to adjust the model to a desired
context. This has been particularly dramatic for molecular solids
which backbones are often so flexible that chemical bonds must
continuously relax or contract to avoid bond breakings. To
overcome these shortcomings, the Morse and double-Morse(DM)
potentials were considered. Thus, the DM potential has been used
in the recent studies of the protonic conductivity~\cite{flytzanis} and the proton ordering
transitions~\cite{tanaka}. For this last context, it has even been
argued~\cite{tanaka} that the DM model is to date the best
candidate giving relatively good account of the so-called
"geometric effect" . \\
 In addition to the DM, several other parametrized DWP models are
 currently present in the literature~\cite{behera,dika1}. By their
 essential virtues they allow theoretical manipulations at one wish.
 The present work deals with two versions among the most general ones.
 A main question in the theory of parametrized DWP (PDWP) is their
 analytical tractability. Indeed, except the soliton solutions that
 are often easy to obtain, several other physical quantities are
 explicitely unaccessible. Even their statistical mechanics are most
 often proceeded numerically~\cite{habib}, or by transfer integral methods
 which however can open only very narrow paths in the rather wide range of
 thermodynamic parameters of the system~\cite{dika2}. For the present two
 variants we will discuss the lattice discreteness properties to both analytical and numerical viewpoints.
 We start by deriving a generalized analytical expression of the Peierls-Nabarro
 potential(PNP)~\cite{bak1}. Next, we carry out
 a numerical analysis to get into the intimate discrete structure of each
 model. By this last way we also provide an adequate framework to
 check the exactness of our analytical results. \\
   Starting, consider a $1D$ nonlinear Klein-Gordon system whose dynamics
   can be described by the Hamiltonian:
\begin{equation}
H= M_o \sum_{n=1} \left[\frac{1}{2}\phi^2_{n,t} + \frac{C^2_o}{2\ell^2} \left(\phi_{n+1}-\phi_n \right)^2 + \omega^2_o V(\phi_{n},\mu)\right] \label{a}
\end{equation}
$V(\phi,\mu)$ is the PDWP assumed of the form~\cite{dika1}:
\begin{equation}
V_{\mu}(\phi) = \frac{\it{a}}{8}\left(\frac{1}{\mu^2} \sinh^2(\alpha \phi) - 1\right)^2 \label{b}
\end{equation}
The quantities $\it{a}$ and $\alpha$ can take different values,
namely:
\begin{equation}
 \it{a}= a_o, \hspace{.2in} \alpha = \mu  \label{c}
\end{equation}
For this value set, (\ref{b}) is a DWP with variable positions of degenerate minima but fixed potential hump.
\begin{equation}
 \it{a}= \frac{a_o \mu^2}{(1 + \mu^2)arsinh^2(\mu)}, \hspace{.2in} \alpha = arsinh(\mu)  \label{d}
\end{equation}
for this second value set (\ref{b}) is a DWP with fixed minima but
variable potential hump. In the continuum limit($x= n\ell$, where
$\ell$ is the lattice spacing), the PDWP admits single-kink
solutions explicitly given by:

\begin{equation}
\phi_n(s=n\ell-vt, \mu)= \pm \frac{1}{\mu}artanh\left[\frac{\mu}{\sqrt{1 + \mu^2}}\tanh \frac{\gamma s}{\sqrt{2}d(\mu)} \right] \label{e}
\end{equation}
Where we set $\gamma^{-2}= 1- \frac{v^2}{C^2_o}$ and call $d(\mu)$
the kink width:
\begin{equation}
d^2(\mu)= \frac{a_o d^2_k}{\it{a}\alpha^2}\frac{\mu^2}{1+\mu^2}, \hspace{.2in} d^2_k = \frac{C^2_o}{\omega^2_o a_o} \label{f}
\end{equation}
$d_k$ in this last relation is the kink "bare" width. At this step
it is instructive to remark that according to the two model
parameters, only the first will give rise to kink family whose
widths effectively vary(decrease) as function of (with increasing)$\mu$, whereas all kinks of
the second family have their widths bounded to the constant $d_k$ whatever $\mu$. However, their "kink shape" is deeply
affected by the deformability parameter as noticed elsewhere~\cite{dika1}. \\
As our primary goal is the kink-lattice interactions, proceeding
with we appeal to the PNP approach which assumes the existence of
an effective potential field provided by the lattice substrate and
via which kinks get pinned due to lattice discreteness.
Several techniques are known for this problem. According to the
recent~\cite{willis}(the collective-coordinate method), lattice
discreteness act on the kink in two different aspects namely on
the kink trajectory(time dependence of the kink center-of-mass
coordinate within the discrete PNP), and shape. The
renormalization of the kink shape(discreteness effect-induced kink
shape dressing) is though to be manifest of radiations by the
pinned kink, of discrete phonons as a shape response to the
lattice discrete structure. To our knowledge there is yet a sound
experimental evidence of such phenomenon, so we will not consider
this aspect. Moreover, we are interested in the kink static
configuration which is a suitable approximation of the
"pinned-kink" configuration so regular in structural transition
processes. Thus we are left with the sole problem of calculating
the PNP. We start by writing the total energy of the discrete
system considering the static part of eq. (\ref{a}):
\begin{equation}
U= \sum_{n=1}G(n), \hspace{.1in} G(n)= M_oC^2_o \left(\frac{d\phi_n}{dn}\right)^2    \label{h}
\end{equation}
The discrete sum in (\ref{h}) can be evaluated using the Poisson sommation formula~\cite{bak1,pokrovsky},
\begin{equation}
U = M_oC^2_o \sum^{\infty}_{m=-\infty} \int^{\infty}_{-\infty} dn \left(\frac{d\phi_n}{dn}\right)^2 \exp{(2\pi imn)} \label{i}
\end{equation}
The integrand in this last equation is derived from the single-kink solution eq. (\ref{e}), which gives:
\begin{equation}
\left(\frac{d\phi_n}{dn}\right)^2 = \left(\frac{B_o}{\cosh(\frac{\sqrt{2}\ell n}{d(\mu)})+ \cosh(X_o)}\right)^2,  \hspace{.1in} B^2_o = \frac{2\ell (1+\mu^2)}{d^2(\mu)} \label{j}
\end{equation}
We take only the real part of eq. (\ref{i}) which is equivalent to restrict the sum over cosine harmonics.
Otherwise, this is consistent with the fact that $G(n)$ is an even function of $n$. It turns out that the
coefficients of this series must be determined by the following class of integrals:
\begin{equation}
\int^{\infty}_0 dX \frac{\cos(XY_m)}{\left[\cosh(X)+\cosh(X_o) \right]^2} = \sqrt{\frac{\pi}{2}} \frac{\sinh^{3/2}(X_o)\Gamma(2+iY_m)\Gamma(2-iY_m)}{\Gamma(2)} \wp^{-3/2}_{-1/2+iY_m}[\coth(X_o)] \label{k}
\end{equation}
\begin{equation}
X= \frac{2n\ell}{\sqrt{2}d(\mu)}, \hspace{.2in}  Y_m= \frac{ \sqrt{2} \pi m d(\mu)}{\ell} \label{l}
\end{equation}
where $\wp$ is the Associate Legendre
function~\cite{handbook}. The complex-argument Gamma function
$\Gamma$ appearing in this set of integrals is defined such that:
\begin{equation}
\Gamma(2+iY_m)\Gamma(2-iY_m)= \frac{\mid \Gamma(2) \mid^2}{\prod^{\infty}_{k=0}{\left[1+\frac{Y^2_m}{(2+k)^2}\right]}} \label{n}
\end{equation}
The product function is evaluated by the standard formula~\cite{handbook1}:
\begin{equation} 
\prod^{\infty}_{k=0} \left[1+\frac{4Y^2_m}{(2k+1)^2\pi^2} \right] = \cosh(Y_m) \label{o}
\end{equation}
Rewriting (\ref{i}) as a Fourier series in $m$, combinations of  eqs. (\ref{j})-(\ref{o}) lead to the following expression of the Fourier components:
\begin{equation}
U_m = \frac{2 d(\mu) \sqrt{\pi} \sinh^{3/2}(X_o)B^2_oM_oC^2_o}{\ell \cosh(\frac{\sqrt{2} \pi^2 m d(\mu)}{\ell})} \sqrt{\frac{\pi}{2}}\wp^{-3/2}_{-1/2+iY_m}[\coth(X_o)] \label{r}
\end{equation}
The most indicated framework where to check the consistency of our
analytical theory is the numerical treatment. We performed the
integration in (\ref{i}) following an extended midpoint scheme.
Results are displayed on figures 1 and 2 for arbitrary values of
the deformability parameter $\mu$. The main variable here is the
dimensionless kink bare width $\frac{d_k}{\ell}$. $d_k$
corresponds to the asymptotic limit $\mu \rightarrow 0$, where the
two models reduce to the well-known $\phi^4$ model. Figures 1a and
1b are the kink continuum energies at rest i.e. $E_k=Uo$, while
figures 2a and 2b represent the PNP barriers $E_{PN}= U_1$. Curves
are plotted in unit of $M_oC^2_o$. Though the figures indicate
rather complex behaviours of the PNP barrier with respect to
$\mu$, their most stricking feature is the drastic exponential
fall-off of the pinning barrier with increasing $\mu$. For the
first model, this feature relates to the fact that the greater the
deformability parameter the narrower the kink. On the contrary, the kink
width in the second model is independent on $\mu$ but however, one clearly
sees that the corresponding PNP barriers show almost the same
dependence on $\mu$ as in the first model. \\
The drastic fall-off of the PNP barriers can be interpreted in terms of the kinks
 the overcoming lattice pinning effect. Such a behaviour suggests an enhancement of the 
conducting regime in the discrete system at relatively narrow kink
widths. This sounds consistent with previous
predictions~\cite{jensen,bak1,pokrovsky,willis} within the $\phi^4$ theory. otherwise, by tuning the
deformability parameter(which in the context of 1D molecular
systems may correspond to adjust the equilibrium positions of
ions, for instance), it is expected that the correct order of magnitude of
transport parameters of DWP systems will be obtained with
relatively best accuracy~\cite{woafo}.

\begin{acknowledgments}

The author wishes to thank P. Woafo and T. C. Kofan\'e of the University of yaound\'e for enriching discussions.

\end{acknowledgments}

\newpage

\section{Figure Captions:}

Figure 1a: Reduced continuum energy of a kink at rest for the
first PDWP model in the text. \\

Figure 1b: Reduced continuum energy of a kink at rest for the
second PDWP model. \\

Figure 2a: The reduced Peierls-Nabarro potential amplitudes for
the first PDWP model.\\

 figure 2a: The reduced Peierls-Nabarro potential amplitudes for the second PDWP model.

\end{document}